\documentclass[sigconf]{acmart}

\usepackage[english]{babel}
\usepackage[]{color}
\usepackage{listings}
\usepackage{xcolor}
\usepackage{colortbl}
\usepackage{comment}
\usepackage{graphicx}
\usepackage{epsfig}
\usepackage{array, colortbl}
\usepackage{listings}
\usepackage{epstopdf}
\usepackage{multirow}
\usepackage{rotating}
\usepackage{float}
\usepackage[]{url}
\usepackage{balance}
\usepackage{fancybox}
\usepackage{scalefnt}
\usepackage[normalem]{ulem}
\usepackage{subcaption}
\usepackage{cleveref}
\usepackage{xspace}
\usepackage{caption}
\usepackage{changepage}
\usepackage{booktabs}
\usepackage{diagbox}
\usepackage{natbib}
\usepackage{tabularx}
\usepackage{enumitem}
\setlist[enumerate]{leftmargin=*}

\definecolor{amber}{rgb}{1.0, 0.49, 0.0}
\definecolor{lavenderindigo}{rgb}{0.58, 0.34, 0.92}
\definecolor{islamicgreen}{rgb}{0.0, 0.56, 0.0}

\definecolor{LightCyan}{rgb}{0.88,1,1}
\definecolor{beaublue}{rgb}{0.74, 0.83, 0.9}
\definecolor{bubbles}{rgb}{0.91, 1.0, 1.0}

\newcommand{\npm}{\texttt{npm}\xspace}
\newcommand{\npminstall}{\texttt{npm install}\xspace}
\newcommand{\packagejson}{\texttt{package.json}\xspace}
\newcommand{\nodemodules}{\texttt{node\_modules}\xspace}

\newcommand{\webpack}{\texttt{webpack}\xspace}
\newcommand{\rollup}{\texttt{rollup}\xspace}

\newcommand{\conclusion}[1]{\begin{center}\begin{tcolorbox}[skin=widget, left=0.9mm,right=0.9mm,top=0.9mm,bottom=0.9mm,boxrule=0.3mm,arc=0mm,coltitle=black,colframe=black!99!white,colback=white!88!gray,width=(\linewidth),before=\hfill,after=\hfill]#1\end{tcolorbox}\end{center}}

\newcommand{\ra}[1]{\renewcommand{\arraystretch}{#1}}

\newcommand{\rqi}{How many installed dependencies are production dependencies?}
\newcommand{\rqii}{What are the characteristics of production dependencies?}
\newcommand{\rqiii}{How often are npm security alerts emitted for production dependencies?}

\definecolor{ScarletRed}{rgb}{0.80,0.00,0.00}

\newcommand{\diego}[1]{}%

\makeatletter
\newcommand{\labelname}[1]{%
  \def\@currentlabelname{#1}}%
\makeatother

\usepackage[skins]{tcolorbox}

\colorlet{punct}{red!60!black}
\definecolor{background}{HTML}{EEEEEE}
\definecolor{delim}{RGB}{20,105,176}
\colorlet{numb}{magenta!60!black}

\lstdefinelanguage[ECMAScript2015]{JavaScript}[]{JavaScript}{
  morekeywords=[1]{await, async, case, catch, class, const, default, do,
    enum, export, extends, finally, from, implements, import, instanceof,
    let, static, super, switch, throw, try},
  morestring=[b]` %
}

\lstdefinelanguage{JavaScript}{
  morekeywords=[1]{break, continue, delete, else, for, function, if, in,
    new, return, this, typeof, var, void, while, with, dependencies, devDependencies},
  morekeywords=[2]{false, null, true, boolean, number, undefined,
    Array, Boolean, Date, Math, Number, String, Object, render},
  morekeywords=[3]{eval, parseInt, parseFloat, escape, unescape},
  sensitive,
  morecomment=[s]{/*}{*/},
  morecomment=[l]//,
  morecomment=[s]{/**}{*/}, %
  morestring=[b]',
  morestring=[b]"
}[keywords, comments, strings]

\lstalias[]{ES6}[ECMAScript2015]{JavaScript}

\definecolor{mediumgray}{rgb}{0.3, 0.4, 0.4}
\definecolor{mediumblue}{rgb}{0.0, 0.0, 0.8}
\definecolor{forestgreen}{rgb}{0.13, 0.55, 0.13}
\definecolor{darkviolet}{rgb}{0.58, 0.0, 0.83}
\definecolor{royalblue}{rgb}{0.25, 0.41, 0.88}
\definecolor{crimson}{rgb}{0.86, 0.8, 0.24}

\lstdefinestyle{JSES6Base}{
  backgroundcolor=\color{background},
  basicstyle=\ttfamily,
  breakatwhitespace=false,
  breaklines=false,
  captionpos=b,
  columns=fullflexible,
  commentstyle=\color{mediumgray}\upshape,
  emph={},
  emphstyle=\color{crimson},
  extendedchars=true,  %
  fontadjust=true,
  frame=single,
  identifierstyle=\color{black},
  keepspaces=true,
  keywordstyle=\color{mediumblue},
  keywordstyle={[2]\color{darkviolet}},
  keywordstyle={[3]\color{royalblue}},
  numbers=left,
  numbersep=5pt,
  numberstyle=\tiny\color{black},
  rulecolor=\color{black},
  showlines=true,
  showspaces=false,
  showstringspaces=false,
  showtabs=false,
  stringstyle=\color{forestgreen},
  tabsize=2,
  title=\lstname,
  upquote=true  %
}

\lstdefinestyle{JavaScript}{
  language=JavaScript,
  style=JSES6Base
}
\lstdefinestyle{ES6}{
  language=ES6,
  style=JSES6Base
}

\lstdefinelanguage{json}{
    basicstyle=\ttfamily,
    numbers=left,
    numberstyle=\tiny\color{black},
    captionpos=b,
    columns=fullflexible,
    numbersep=5pt,
    showstringspaces=false,
    fontadjust=true
    breaklines=False,
    frame=single,
    backgroundcolor=\color{background},
    morekeywords=[1]{dependencies, devDependencies, version, file, mappings, sources, sourcesContent},
    keywordstyle=\color{mediumblue},
    literate=
     *{0}{{{\color{numb}0}}}{1}
      {1}{{{\color{numb}1}}}{1}
      {2}{{{\color{numb}2}}}{1}
      {3}{{{\color{numb}3}}}{1}
      {4}{{{\color{numb}4}}}{1}
      {5}{{{\color{numb}5}}}{1}
      {6}{{{\color{numb}6}}}{1}
      {7}{{{\color{numb}7}}}{1}
      {8}{{{\color{numb}8}}}{1}
      {9}{{{\color{numb}9}}}{1}
      {:}{{{\color{punct}{:}}}}{1}
      {,}{{{\color{punct}{,}}}}{1}
      {\{}{{{\color{delim}{\{}}}}{1}
      {\}}{{{\color{delim}{\}}}}}{1}
      {[}{{{\color{delim}{[}}}}{1}
      {]}{{{\color{delim}{]}}}}{1},
}

\AtBeginDocument{%
  \providecommand\BibTeX{{%
    \normalfont B\kern-0.5em{\scshape i\kern-0.25em b}\kern-0.8em\TeX}}}

\copyrightyear{2022}
\acmYear{2022}
\setcopyright{acmcopyright}\acmConference[ASE '22]{37th IEEE/ACM International Conference on Automated Software Engineering}{October 10--14, 2022}{Rochester, MI, USA}
\acmBooktitle{37th IEEE/ACM International Conference on Automated Software Engineering (ASE '22), October 10--14, 2022, Rochester, MI, USA}
\acmPrice{15.00}
\acmDOI{10.1145/3551349.3556896}
\acmISBN{978-1-4503-9475-8/22/10}

\begin{document}

\title{Not All Dependencies are Equal: An Empirical Study on Production Dependencies in NPM}

\author{Jasmine Latendresse}
\affiliation{%
\institution{Data-driven Analysis of Software (DAS) Lab \\ Concordia University}
\city{Montreal}
\country{Canada}}
\email{jasmine.latendresse@concordia.ca}

\author{Suhaib Mujahid}
\affiliation{%
\institution{Mozilla Corporation}
\city{San Francisco}
\country{United States}}
\email{smujahid@mozilla.com}

\author{Diego Elias Costa}
\affiliation{%
\institution{LATECE Lab \\ Université du Québec à Montréal}
\city{Montreal}
\country{Canada}}
\email{costa.diego@uqam.ca}

\author{Emad Shihab}
\affiliation{%
\institution{Data-driven Analysis of Software (DAS) Lab \\ Concordia University}
\city{Montreal}
\country{Canada}}
\email{emad.shihab@concordia.ca}

\begin{abstract}
Modern software systems are often built by leveraging code written by others in the form of libraries and packages to accelerate their development. While there are many benefits to using third-party packages, software projects often become dependent on a large number of software packages. Consequently, developers are faced with the difficult challenge of maintaining their project dependencies by keeping them up-to-date and free of security vulnerabilities. However, how often are project dependencies used in production where they could pose a threat to their project's security?

We conduct an empirical study on 100 JavaScript projects using the Node Package Manager (npm) to quantify how often project dependencies are released to production and analyze their characteristics and their impact on security. 
Our results indicate that less than 1\% of the installed dependencies are released to production. Our analysis reveals that the functionality of a package is not enough to determine if it will be released to production or not. In fact, 59\% of the installed dependencies configured as runtime dependencies are not used in production, and 28.2\% of the dependencies configured as development dependencies are used in production, debunking two common assumptions of dependency management. Findings also indicate that most security alerts target dependencies not used in production, making them highly unlikely to be a risk for the security of the software. Our study unveils a more complex side of dependency management: not all dependencies are equal. Dependencies used in production are more sensitive to security exposure and should be prioritized. However, current tools lack the appropriate support in identifying production dependencies. 
\end{abstract}


\maketitle

\section{Introduction}
\label{sec:introduction}
The vast majority of modern software systems are built by using modular functionalities provided by open source packages. 
Reports estimate that more than 90\% of open source and proprietary projects rely substantially on reusing open source packages~\cite{github_octoverse,a2019_eight}.  
As a testament to the popularity of open source, popular package managers such as \npm, host more than 2 million reusable packages, covering all sorts of software functionalities~\cite{decan}.

While the use of open source packages significantly reduces development time and costs~\cite{inbook, murphyhill_2019_what, basili_1996_how}, it also exposes software applications to vulnerabilities. 
In the 2020 State of the Octoverse security report, GitHub reveals that active repositories with a supported package ecosystem have a 59\% chance of getting a security alert in the next 12 months \cite{github_octoverse}. 
This problem is even more widespread in the JavaScript ecosystem, where nearly 40\% of all \npm packages rely on code with known vulnerabilities~\cite{sonatype}.  
Software vulnerabilities may lead to significant financial and reputation loss. A popular example is the 2017 Equifax cybersecurity incident caused by a web-server vulnerability in the Apache Struts package. The incident led to a data breach of millions of American citizens, costing Equifax 1.8 billion USD in security upgrades and lawsuits~\cite{fruhlinger_2020_equifax}.

The problem is that developers struggle to identify what vulnerabilities may affect their software application~\cite{Kula_2017}.
Current security scanners report on the severity of a vulnerability, but lack a support to identify if the dependency is 1) used in the code and 2) is part of the production software the project delivers. 
Developers constantly complain that security alert tools report too many false positives~\cite{pashchenko_2018_vulnerable,pashchenko}, as even the most critical vulnerability may be unexploitable if the vulnerable dependency is never released in the production software.

In this paper, we study how often dependencies are actually part of a production software and their impact on security based on their characteristics, usage, and context. 
We perform this study on 100 JavaScript projects in \npm, the largest and fastest growing software ecosystem to date~\cite{stackoverflow:21:survey}, to answer the following three research questions: 

\begin{table*}[t]
\caption{Concepts and definitions. 
\label{tab:concepts_and_defs}}
\small
\newcommand{\grayrow}{\rowcolor{gray!10}}

\begin{tabular}{lll}
\toprule
\textbf{Concept}     & \textbf{Definition}    & \textbf{Example} \\
\midrule
\textbf{Runtime dependency}     & Refers to the "runtime" configuration of a dependency  & \texttt{react} is a runtime dependency as shown in Figure \ref{fig:packagejson}.\\
& in the \packagejson file and is needed for the  & \\
& application to function. & \\
\grayrow
\textbf{Development dependency} & Refers to the "development" configuration of a dependency  & \texttt{webpack} is a development dependency as shown in \\
\grayrow
& in the \packagejson file and indicates that the & Figure \ref{fig:packagejson}.\\
\grayrow
& dependency is needed to develop the application. & \\
\textbf{Installed dependency} & Refers to the dependencies installed in the project & The dependencies depicted in Figure \ref{fig:packagejson} are part of the \\
& and the result of the \npminstall command. & installed dependencies, and so are their dependencies. \\
\grayrow
\textbf{Depth}      & Refers to the level of a dependency in the dependency tree. &  \texttt{npm ls} is used to obtain the dependency tree. \\
\textbf{Direct dependency} & Refers to a dependency with a depth of 1. & The dependencies shown in Figure \ref{fig:packagejson} are direct \\
& & dependencies. \\
\grayrow
\textbf{Transitive dependency} & Refers to a dependency with a depth greater than 1. & The dependencies of the dependencies shown in \\
\grayrow
& & Figure \ref{fig:packagejson} are transitive dependencies. \\
\midrule
\textbf{Usage}       & Refers to the scope in which a dependency is used & Figure \ref{fig:sourcemap} shows that \texttt{react-dom} is used in production. \\
\grayrow
\textbf{Context}     & Refers to the context of the application in  & In our example application, \texttt{webpack} is a development  \\
\grayrow 
& which a dependency is used. & dependency used to bundle the application's resources. \\

\bottomrule                   
\end{tabular}

\end{table*}

\begin{enumerate}[align=left]
    \item[RQ1.] \rqi
    \item[RQ2.] \rqii
    \item[RQ3.] \rqiii
\end{enumerate}

Findings show that production dependencies represent a very small fraction of the total number of dependencies in each project. 
While projects tend to depend on hundreds of dependencies (both direct and transitive), 51 projects did not have any production dependencies, and 49 have a median of 5 production dependencies. 
Contrary to common assumptions, most dependencies declared as runtime are not shipped to production while some development dependencies are included in the production software.
Consequently, we find that dependency usage and context gives better insight at determine if a dependency will be used in production than the nature of a dependency itself. Furthermore, our results show that not all security vulnerabilities reported by \npm are an actual threat to the software in production. Our paper makes the following contributions:

\begin{itemize}
    \item To the best of our knowledge, this is the first study to investigate the discrepancy between installed and production dependencies in open source projects.
    \item We report on results that challenge the assumptions of dependency management and should be revisited by researchers and practitioners.  
    \item We investigate the support of current tools in providing better information for developers regarding the scope and context of vulnerable dependencies.
    \item We make our dataset of 100 projects available\footnote{\url{https://zenodo.org/record/6518765}}, including all scripts used to collect and pre-process data, to facilitate replication and foment more research in the field. 
\end{itemize}

The rest of the paper is organized as follows: we start by motivating our problem with an example in Section \ref{sec:background}. 
We describe and justify our methodology in Section \ref{sec:design} and explain our results in Section \ref{sec:results}. Implications of our findings are discussed in Section \ref{sec:discussion}. We present the related work in Section \ref{sec:related_work}, and discuss the limitations to our study in Section \ref{sec:threats_to_validity}. Finally, we conclude our study in Section~\ref{sec:conclusion}.

\section{Motivation \& Background}
\label{sec:background}

To motivate our study and illustrate the terminology used in this paper, we walk the reader through the creation of a simple application using \texttt{create-react-app}~\cite{create-react-app}. The terms used in this example and throughout this paper are formally defined in Table \ref{tab:concepts_and_defs}.
This example application is a single-page "Hello World" application that is provided by React when initializing a Create React App project. We create our application by simply running the command \texttt{npm create-react-app my-app}. 

\textbf{How many dependencies in our project?}
To achieve this single-page React application without further programming, our generated application reuses several open source packages published in \npm. 
We refer to each of the packages as a \textit{dependency} of our project. 
The dependency configuration of our project is stored in the \packagejson file, shown in  Figure~\ref{fig:packagejson}. 
Dependencies are grouped into two groups: \textit{runtime dependencies} (``dependencies'') and \textit{development dependencies} (``devDependencies'').
Runtime dependencies are dependencies required by the application to function, e.g., as we build a React application, our project depends on \texttt{react} version 17.0.2. 
Development dependencies, on the other hand, are needed to develop the project, e.g., to format the code (prettier 2.5.1), and are not required by the software to run. 
As it can be seen in Figure~\ref{fig:packagejson}, our small application has 7 runtime dependencies and 9 development dependencies. 

\begin{figure}
\vspace{-.4cm}
    \caption{A snippet of the \packagejson file listing the dependencies of our example project.}
    \label{fig:packagejson}
    \begin{lstlisting}[language=json,firstnumber=1,frame=single,basicstyle=\small]
    "dependencies": {
        "@testing-library/jest-dom": "^5.16.2",
        "@testing-library/react": "^12.1.3",
        "@testing-library/user-event": "^13.5.0",
        "react": "^17.0.2",
        "react-dom": "^17.0.2",
        "react-scripts": "5.0.0",
        "web-vitals": "^2.1.4"
      },
      "devDependencies": {
        "@webpack-cli/generators": "^2.4.2",
        "css-loader": "^6.6.0",
        "html-webpack-plugin": "^5.5.0",
        "prettier": "^2.5.1",
        "style-loader": "^3.3.1",
        "webpack": "^5.69.1",
        "webpack-cli": "^4.9.2",
        "webpack-dev-server": "^4.7.4",
        "workbox-webpack-plugin": "^6.5.0"
      }
    \end{lstlisting}
    \vspace{-.4cm}
\end{figure}

Once we install these dependencies locally to build and test our application (\texttt{npm install}) we may be surprised to see that a total of 1,764 dependencies were installed.
The dependencies shown in Figure~\ref{fig:packagejson} are \textit{direct dependencies} of our project, each of which have dependencies of their own. 
For instance, the package \texttt{loose-envify} is a dependency of \texttt{react}.
These are called \textit{transitive dependencies} and represent the vast majority of installed dependencies. As such, \texttt{loose-envify} is a transitive dependency of our example application.
We use the term \textit{installed dependencies} to refer to all dependencies of a project, both direct/transitive and development/runtime dependencies.

\textbf{Is our application vulnerable?}
Security vulnerabilities are a widespread problem in \npm~\cite{sonatype}. Given that our application depends on 1,764 installed dependencies, is our application affected by vulnerabilities?
To verify this, we resort to using a Software Composition Analysis (SCA) tool.
SCA tools are used to identify open source components in software codebases to evaluate security, license compliance and overall code quality \cite{sca_def}.
In our example, we use \textit{npm audit}, a native tool of \npm that reports vulnerabilities affecting software dependencies and maintains its own database of vulnerabilities. 
If a dependency is affected by one of more vulnerabilities, we refer to the dependency as a \textit{vulnerable dependency}.
In our example application, upon running \texttt{npm audit}, we receive the report that our simple application contains 6 moderate severity vulnerabilities, 13 high severity vulnerabilities, and 1 critical severity vulnerability.
That is, without any further programming, our project already started with an alarming number of vulnerabilities of moderate, high, and critical severity. Examples of the reported high severity and critical severity vulnerabilities include Regular Expression Denial of Service, Template Injection, and Prototype Pollution.

\textbf{Can reported vulnerabilities really affect our example application in production?}
Vulnerable dependencies are problematic and may affect the security of our project in multiple ways. 
However, the risk of vulnerable dependencies reaches its peak when the dependency is needed for the software to run in a production environment. 
To find which dependencies are part of our production software, i.e., \textit{production dependencies}, we use a module bundler. 
A \textit{module bundler} is a tool that assists the building process of a software by resolving the software dependencies and pruning the dependencies that are not needed in the production software. The process of pruning dependencies is referred to as \textit{tree shaking}. 
We use \textit{webpack} \cite{a2019_webpack}, a popular JavaScript module bundler, to build our production software and export a list of production dependencies. 

Upon building our project with \webpack, the tool generates a \textit{source map} file, which contains the list of production dependencies of our application. 
From the 1,764 dependencies in our example project, Figure \ref{fig:sourcemap} shows that only 6 are released to production: \texttt{react}, \texttt{object-assign}, \texttt{scheduler}, \texttt{react-dom}, \texttt{style-loader}, and \texttt{css-} \texttt{loader}. 
More so, none of our production dependencies contained any reported vulnerability, thus, our original report of 15 vulnerabilities affected dependencies that would not be present in the application in production. 

\textbf{The problem: security alert fatigue.} 
Our example showcases an important problem in current software development.
Even small applications may depend on thousands of dependencies and vulnerabilities are constantly being reported by the open source community.
Developers face the difficult challenge of separating security alerts that are relevant to their application security from the long reports yielded by current SCA tools~\cite{pashchenko_2018_vulnerable,pashchenko}. 
In this paper, we evaluate this problem on a scale of 100 popular JavaScript projects.

\begin{figure}
\vspace{-.4cm}
    \caption{A snippet of the source map generated by building our example project.}
    \label{fig:sourcemap}
    \begin{lstlisting}[language=json,firstnumber=1,frame=single,basicstyle=\small]
    "version": 3,
    "file": "main.js"
    "mappings": "KAAK,CAACC,EAAOC,GAAI..."
    "sources": ["node_modules/css-loader/dist/runtime/api.js",
    "node_modules/object-assign/index.js",
    "node_modules/react-dom/cjs/react-dom.production.min.js",
    "node_modules/react/cjs/react.production.min.js",
    "node_modules/scheduler/index.js",
    "node_modules/style-loader/injectStylesIntoStyleTag.js"]
    \end{lstlisting}
    \vspace{-.4cm}
\end{figure}

\section{Study Design}
\label{sec:design}

\begin{figure*}
    \centering
    \includegraphics[scale=0.65]{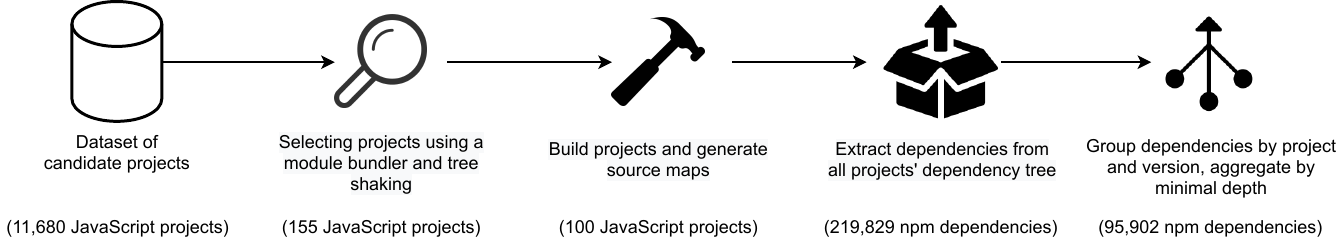}
    \caption{Overview of our approach for filtering projects and collecting dependencies.}
    \label{fig:approach}
\end{figure*}

The goal of the paper is to study how often project dependencies are shipped to production and their impact on the security of software projects.  
In this section, we describe how we select and curate the set of study projects (Sections~\ref{sec:candidate} and~\ref{sec:pruning}), and how we identify production dependencies (Section~\ref{sec:identifydeps}).
We provide an overview of our methodology in Figure \ref{fig:approach}.

\subsection{Dataset of Candidate Projects}
\label{sec:candidate}

The focus of our study is to investigate how active software development JavaScript projects use their dependencies. To this aim, we start by collecting data of a large number of JavaScript repositories as candidate projects for our study. Many studies have used the number of GitHub stargazers as a way to select candidate projects~\cite{ray,guzman,Costa:21:GO}. 
Thus, we start with 11,860 popular JavaScript projects that were collected on July 27th, 2020 with at least 100 stargazers.

Finding what dependencies are shipped to production is a very challenging task making it impractical to apply this analysis on a large-scale \cite{zapata}. 
In our study, we opt to select projects that already make use of tree shaking (see Section~\ref{sec:background} for a more in-depth explanation). 
Specifically, we select projects using either \webpack or \rollup \cite{rollupjs} because they are two of the most popular module bundlers for JavaScript projects and they have integrated tree shaking support.

To find out which projects use \webpack and \rollup, we automatically parse the \packagejson files of the 11,860 projects to identify 1) if any of the bundlers are declared as a dependency and 2) the tree shaking algorithm is enabled for the project. 
Through this process, we find that 155 JavaScript projects make use of \webpack or \rollup, and have tree shaking enabled.

\begin{table}[b]
\caption{Descriptive Statistics of the Selected Projects.\label{tab:dataset_stats}}
\centering
\begin{tabular}{rrrrr}
\toprule
                          & \textbf{Mean} & \textbf{Median} & \textbf{Min} & \textbf{Max} \\
                          \midrule
\textbf{\# stars}            & 4827.6        & 1224            & 112          & 74201        \\
\textbf{\# commits}      & 1364.9        & 496             & 31           & 6188         \\
\textbf{\# contributors} & 62.3          & 26              & 4            & 401          \\
\textbf{age (years)}              & 5.2           & 5               & 1            & 12           \\ \bottomrule
\end{tabular}
\end{table}

\subsection{Building Candidate Projects}
\label{sec:pruning}

To assess whether a dependency is used in production, we have to successfully build each project in a production environment with a module bundler.  
During the build of a project, the module bundler first looks for all of the dependencies in the project and constructs a dependency graph (dependency resolution). The dependency graph is then converted along with source code into a single file (packing) called the bundle. 
Source maps are then generated after a successful build. 

To build the candidate projects, we first clone each of the 155 JavaScript projects locally. 
We planned to build a framework to automate the build of all the 155 JavaScript projects. 
However, we soon realized that many projects require specific building commands and setup to be build successfully.
In fact, the majority of the projects did not support the standard build command (\texttt{npm run-script build}). 
Furthermore, the environmental settings varied across projects, e.g., some projects require specific NodeJS versions and identifying this automatically is very challenging. 
We then proceed to semi-manually build each project using the following methodology:

\begin{enumerate}
  \item \textbf{Read projects documentation.} 
  We read the documentation of all the 100 studied projects to identify the specifics of each project build. The goal of this step is to identify all the steps of the building process: the build commands, supported Node versions, supported package manager (e.g. YAML or npm), and any other specificity of the project building configuration. 
  At this point, we also confirmed that all selected projects are related to software development, i.e., are not personal toy-projects.  
  
  \item \textbf{Install dependencies.} We install all dependencies specified in the \packagejson file by using the \npminstall command. This generates a \nodemodules folder in every project's home directory which contains all installed dependencies.
  
  \item \textbf{Build project.} Following each project's documentation, we build each project in the dataset. 
  The first author manually followed the steps of the building process to ensure the build was successful, the source maps containing the production dependencies was generated, and the yielded artifacts targeted the production environment.  
  
  \item \textbf{Generate source maps.} Upon the successful completion of the building process, source maps are generated and saved in the project's temporary folder or home directory. 
  
\end{enumerate}

After our careful process, we successfully build and generate source maps for 100 JavaScript projects.
From the 55 projects that failed in our process, the main culprit was the generation of the source maps file. 
In most of the failed cases, projects' configuration did not have the flexibility to generate the source maps file. 
For example, we found some projects created using the \texttt{create-react-app} package that does support module bundlers and tree shaking, but does not have the option to output source maps.

We present descriptive statistics of the 100 projects we successful built and generate source maps in Table \ref{tab:dataset_stats}. 
The projects of our dataset are very popular (median 1,224 stargazers), tend to be mature projects (median of 5 years of development and 496 commits) and are developed by medium-sized team of developers (median of 26 developers).

\subsection{Identifying Dependencies in Production}
\label{sec:identifydeps}
\label{sec:missing_peer}

To identify production dependencies, we first collect all dependencies found in the source maps of each projects using a mix of source map parser \cite{unisil_2021_source} and regular expression (regex). 
Then, to obtain the version of each dependency found in the source maps, we locate its \packagejson file in the respective project's \nodemodules folder and parse it. This results in a dataset of production dependencies with their corresponding version. 

In addition to identifying dependencies used in production, we also want to identify two very important characteristics of all dependencies, as they have an influence on the risk of vulnerabilities~\cite{imtiaz_2021_a}: 
1) the dependency scope, runtime or development and 
2) whether the dependency is a direct or transitive dependency of the project. 
To classify a dependency into runtime or development, we analyse the \packagejson file of a project, classifying dependencies configured in the "dependency" section as runtime dependencies, and classifying dependencies declared in "devDependency" as development dependencies. 
Since transitive dependencies are not listed in the \packagejson file, we identify the type of the original dependency which determines the type of the transitive dependency.  

To classify installed dependencies into direct or transitive dependencies, using the command \texttt{npm list} we generate the \textit{dependency tree}, a hierarchical representation of relationship between dependencies.
The \texttt{npm list} command lists all installed dependencies in json format, including the name, version, path, and depth of each dependency. 
From the depth, we identify each dependency as direct or transitive, i.e., direct dependencies have depth $=$ 1, while transitive dependencies have depth $>$ 1.

Our methodology has one limitation, we cannot automatically resolve missing peer dependencies. 
Peer dependencies are used to decouple dependencies between projects, to ensure a single version of the package is installed for all dependencies.
For example, in applications with many \npm packages depending on \texttt{react}, \texttt{react} can be declared as a peer dependency to prevent the installation of multiple (possibly conflicting) versions of \texttt{react}.
Unlike runtime and development dependencies, peer dependencies are not automatically installed by \npm.
Instead, they must be included by the code that uses the package as a dependency. 
We find that 37 projects in our dataset have missing peer dependencies. Since it is not possible to automatically resolve missing peer dependencies for all 37 projects, we exclude the dependencies from our analysis.

\section{Results}
\label{sec:results}
In this section, we present the results of our three research questions. For each research question, we present its motivation, the approach to answer the question, and the results.

\subsection*{RQ1: \rqi}
\noindent\textbf{Motivation:} 
While reusing packages may reduce development time, developers have to constantly maintain their dependencies to fix bugs in the packages and mitigate the problems of vulnerable dependencies~\cite{atique_2019_on,javanjafari_2021_dependency,alfadel_dependabot}.
However, identifying dependencies used in production is not a trivial task making it difficult to prioritize dependency-related maintenance activities \cite{pashchenko_2018_vulnerable}. 

In this research question, we want to assess how often dependencies of the selected projects are actually production dependencies.
Answering this question is the first step to understand how often a runtime and development dependency is used in production. It will also help us better understand how dependencies are used in practice and how they impact the security of software.

\noindent\textbf{Approach:} 
To approach this research question, we use the methodology described in Section \ref{sec:missing_peer}.
That is, we start by installing all dependencies from each project to retrieve the list of installed dependencies and their respective versions. 
To classify an installed dependency into direct or transitive, we generate the dependency tree of each studied project. 
Then, to identify production dependencies, we build all software projects with their respective module bundler (\webpack or \rollup). 
This building process was done manually by following the building steps specified in the project documentation, to ensure each project is built correctly and without errors. 
After building the project, we analyze the yielded source maps to identify the production dependencies.  
Finally, we cross reference the installed dependencies and production dependencies to classify each project dependency into production/non-production, runtime/development, direct/transitive and report our findings.

\begin{table}
\centering
\caption{Dependency profile of projects with and without production dependencies in absolute numbers and median of aggregated value per project. The percentages are always in relation to the \# of Installed Dependencies. }
\vspace{-.1cm}
\label{tab:dataset_split}
\small
\newcommand{\grayrow}{\rowcolor{gray!10}}

\ra{1.2} %
\begin{tabular}{rrr|rr}
\toprule

    & \multicolumn{2}{c|}{\textbf{Projects with Zero}} & \multicolumn{2}{c}{\textbf{Projects with 1+ }} \\
        & \multicolumn{2}{c|}{\textbf{Production Deps}} & \multicolumn{2}{c}{\textbf{Production Deps}}  \\
    
    \midrule
   \textbf{Dependencies} & \textbf{Total} & \textbf{Median} & \textbf{Total} & \textbf{Median}\\
   
   \midrule

    \textbf{Installed}   &
    46,031 (100\%) & 851 & %
    53,421 (100\%) & 1,017 %
    
    \\
    
    \grayrow
    \textbf{Runtime} & 
    1,005 (2.1\%) & 0 & 
    873 (1.6\%) & 5
    
    \\
    
    \grayrow
    \textbf{Dev}  &
    45,025 (97.9\%) & 832 &
    52,542 (98.4\%) & 1,017 
    
    \\
    
    \textbf{Direct} & 
    1,539 (3.4\%) & 40 &
    2,098 (3.9\%) & 29
    
    \\
    
    \textbf{Transitive} & 
    44,492 (96.6\%) & 809 &
    51,307  (96.1\%) & 963
    
    \\
    
    \grayrow
    \textbf{Production} & 
   -- & -- & 
    497 (0.9\%) & 5 
    
    \\

\bottomrule

\end{tabular}
\end{table}

\noindent\textbf{Finding 1: Of the 100 projects, 51 projects contain no production dependencies.} 
To make a better sense of our results, we split our dataset of 100 projects into two sets: projects with production dependencies (49 projects) and projects without production dependencies (51 projects).
Table \ref{tab:dataset_split} shows the total number of dependencies and their characteristics in both sets of projects. 
The 51 projects with no production dependencies have installed a total of 46 thousand dependencies, including direct and transitive packages, however, none of the installed dependencies are used in production. 
More interestingly, among the installed dependencies, there were 1,005 packages that were declared to be runtime dependencies, which is supposedly required at the runtime of the the final software, but were not included in the final production artifact.   

We also note that the set of 51 projects with no production dependencies have a median of runtime dependencies of zero. 
We confirm this finding through manual investigation and find that 39 projects in our dataset only declare development dependencies. 
Most of these projects are libraries meant to be used by other projects as development tools. 
Examples of such projects are~\texttt{Vuex}, a state management pattern for Vue.js applications; \texttt{three.js}, a popular cross-browser 3D library; and \texttt{polished}, a lightweight toolset for writing styles in Javascript. All those library projects have the incentive to depend on little to no runtime dependencies, as the fewer dependencies they have, the less constrained their users may be to rely on their libraries \cite{abdalkareem_2020_on, chen}.

\noindent\textbf{Finding 2: From the 49 projects with production dependencies, production dependencies represent less than 1\% of the installed dependencies.} 
The results show that projects with production dependencies have a total of 53,421 dependencies, of which only 497 dependencies (0.9\%) are released to production (see Table~\ref{tab:dataset_split}). 
Analyzing the median number of dependencies per project (see Median column in Table~\ref{tab:dataset_split}), we find that projects have in median 5 production dependencies while depending in median over a thousand dependencies. 
However, we notice that not all runtime dependencies are used in production. 
The total number of runtime dependencies installed (873) far exceeds the number of production dependencies (497), indicating that many runtime dependencies may be incorrectly configured or not used in the code. 

\begin{figure}
    \centering
    \includegraphics[width=.9\linewidth]{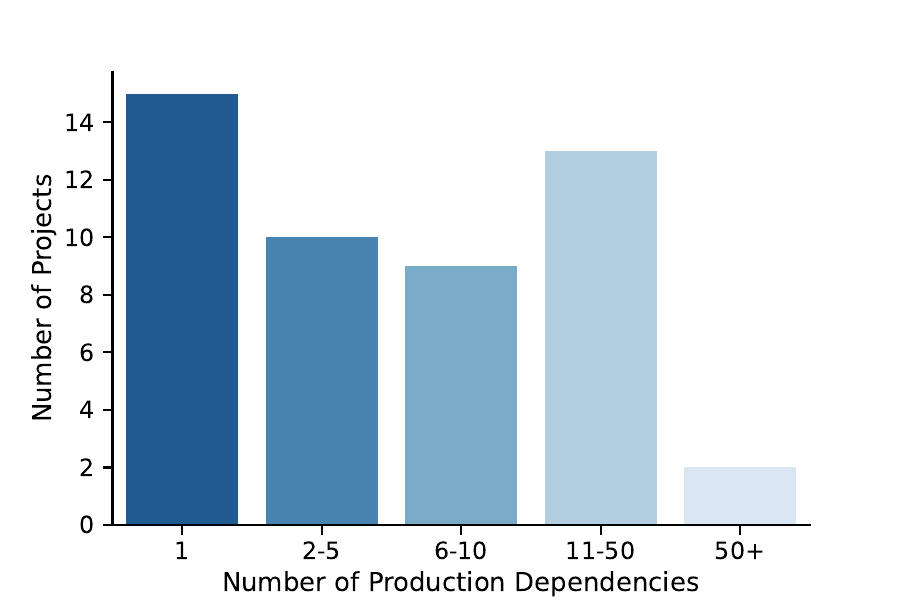}
    \caption{Number of production dependencies on the 49 project with one or more production dependencies. }
    \label{fig:bundled_count}
\end{figure}

Figure \ref{fig:bundled_count} shows the distribution of the number of production dependencies per project. 
We can observe that 15 projects contain a single production dependency and the vast majority of projects (65.3\%) have less than 10 dependencies used in production.
Still, we found some projects that depend heavily on packages in their production build, with \texttt{ProjectMirador\/mirador} being the project with the most production dependencies in our dataset with 92.

\begin{table}[]\centering
\caption{Characteristics of dependencies in projects with production dependencies. 
\vspace{-.1cm}
\label{tab:bundled_characteristics}}
\small
\begin{tabular}{rr rrr}
\toprule
\multicolumn{2}{l}{}                                                        & \multicolumn{1}{c}{\textbf{Direct}} & \multicolumn{1}{c}{\textbf{Transitive}} & \multicolumn{1}{c}{\textbf{Total}}  \\ 
\midrule
\multicolumn{1}{c}{\multirow{2}{*}{\textbf{Production}}}     & \textbf{Dev}     & 62                                & 77                                    & 139                                  \\
\multicolumn{1}{c}{}                                     & \textbf{Runtime} & 175                               & 178                                     & 353                                  \\ \midrule
\multicolumn{1}{c}{\multirow{2}{*}{\textbf{Non-production}}} & \textbf{Dev}     & 1,809                               & 50,594                                   & 52,403              \\
\multicolumn{1}{c}{}                                     & \textbf{Runtime} & 52                                & 458                                     & 510              \\ \midrule
\multicolumn{2}{r}{\textbf{Total}}              & 2,098                               & 51,307                                   & 53,405         \\
\bottomrule
\end{tabular}
\end{table}

\noindent\textbf{Finding 3: More than half (59\%) of the runtime dependencies are not used in production.}
As shown in the "Production" row of Table~\ref{tab:bundled_characteristics}, we find that 510 out of 863 of the total runtime dependencies are not shipped to the production bundle. 
Runtime dependencies are dependencies (supposedly) required by the application to run. 
Our results, however, show that in the majority of the cases, dependencies are declared as runtime but are not actually used in the code, thus, are excluded by the module bundler during the build.
This finding suggests developers mistakenly maintain unused dependencies in their project configuration, which indicates that they lack the necessary information to determine whether a dependency is actually used by the software in production. 
This is corroborated by related work~\cite{javanjafari_2021_dependency}, where authors reported that unused dependencies occur in 80\% of studied projects.

\conclusion{51 out of 100 projects do not use any dependencies in production. 
The 49 projects that ship dependency to production contain less than 1\% of production dependencies. 
Contrary to common belief, 59\% of runtime dependencies are not used in production.
}

\subsection*{RQ2: \rqii}

\noindent\textbf{Motivation:} Production dependencies are the prime security liability in software systems since they can compromise a running software~\cite{zapata}. 
Current SCA tools may not distinguish dependency scope (i.e., production, non-production) \cite{pashchenko_2018_vulnerable, Kula_2017}, which may lead to reporting unexploitable vulnerabilities (false positives). 
They may also only consider direct dependencies although vulnerabilities can be introduced transitively \cite{pashchenko_2018_vulnerable, 7202955, Lauinger_2017}. The problem is that 
assumptions about production dependencies are not always correct. In fact, RQ1 showed that runtime dependencies are not always in production. In this research question, we study the characteristics production dependencies to establish a practical understanding of how they are used and in what context they are used. Such findings help in improving current SCA tools as they provide insights on how dependencies are used in practice.

\noindent\textbf{Approach:} To identify the characteristics of dependencies used in production, we consider the production dependencies identified in RQ1 and classify them based on their scope (runtime, development), depth, and usage.

In theory, one can identify the scope of a dependency by looking at the nature of the functionality provided by a package. 
For instance, packages that provide development utilities should not become production dependencies. 
To investigate to what extent the nature of the package determines if it will be used as a production dependency, we analyze how packages are released to production across the 100 studied projects. 
We analyze a total of 1,269 unique packages. We then classify the packages in three categories: 1) packages that are always used in production, 2) packages that are never used in production, and 3) packages that are sometimes used in production.

\noindent\textbf{Finding 4: 28.2\% of production dependencies are development dependencies.} 
The first section of Table \ref{tab:bundled_characteristics} shows the characteristics of production dependencies. 
We find that 28.2\% of production dependencies are development dependencies and the remaining 71.7\% are declared as runtime dependencies. 
It is expected that all dependencies released to production consist of runtime dependencies since they provide the application with specific functionalities to be used by the client. It is then surprising to find that almost 30\% of the dependencies released to production are development dependencies since such dependencies are, by default, not included in the production bundle.

While unusual, having a development dependency in production occurs in 37 of the projects in our dataset. To better understand this, we perform an exhaustive inspection of the dependency configuration the 37 projects and deduct two possible causes for a development dependency to be in production.
First, the selected projects use module bundlers, which disregard the configuration of the \packagejson file and use source code analysis to identify what should be a production dependency.    
Developers may not be as careful to specify their development dependencies as their building process does not depend on a correct specification of development and runtime dependencies~\cite{a2020_do}. In fact, from the 37 projects with development dependencies in production, 4 (10.8\%) projects declared all of their dependencies as development dependencies although all of them have at least 1 dependency in production. 
Second, it can be that the dependency is initially declared under the "dependencies" property of the \packagejson file, but is intentionally moved by the developer to "devDependencies" to get rid of security warnings, as it is explained in a \texttt{create-react-app} GitHub issue \cite{help}. The author of the issue explains that \texttt{npm audit} reports vulnerabilities for code that never runs in production, but strictly at build time in development. They then suggest to move vulnerable dependencies to "devDependencies" to get rid of the security warning. 
We believe development production dependencies are unlikely to happen in projects that do not use module bundlers. 
By default, \npm does not include development dependencies in a production build.
This means that a project that requires a development dependency at runtime will not function because of the missing dependency.

\noindent\textbf{Finding 5: The majority of production dependencies (51.8\%) are transitive dependencies.} 
Looking at the Transitive columns of Table \ref{tab:bundled_characteristics}, we notice that 51.8\% of production dependencies are dependencies of their direct project dependencies. 
These results suggest that developers may not have control over the majority of production dependencies.
Naturally, a transitive dependency can only be released to production if the original dependency is also released to production. 
Hence, developers should be extra careful when selecting production dependencies, preferably by selecting packages that have little to no production dependencies on their own, to reduce the attack surface through vulnerable dependencies.

\noindent\textbf{Finding 6: The 237 production dependencies come from 183 unique \npm packages. From these, 43 are sometimes not used in production in other projects.}
To put things in perspective, we evaluate the number of unique \npm packages in our dataset by grouping the packages by name and obtain 1,269 unique \npm packages. From this, we find that 1,086 (85.6\%) are never used in production since they offer functionalities that are development-only. For example, \texttt{eslint} installed in 79 projects, is a static code analysis tool that is used to identify problematic patterns found in JavaScript code, \texttt{@babel/core} installed in 70 projects, is a command line interface tool that facilitates working with \texttt{babel}, and \texttt{rollup} installed in 65 projects, is a build tool for JavaScript projects. 

For the rest of the packages, we find that 183 packages are used in production at least once. Taking a closer look at the production packages, we find that 140 (76.5\%) packages are always used in production when installed in a project and that such packages do not occur frequently. In fact, they occur at most in 2 different projects and are installed as runtime dependencies. For example, \texttt{is-promise}, a library that tests whether an object is a \texttt{promises-a+} promise, \texttt{query-string}, a library that parses and stringifies URL query strings, and \texttt{react-fast-compare}, a library that provides specific handling of fast deep equality comparison for React, are all installed in 2 projects, and used in production 100\% of the time they are installed. 

Interestingly, we find that 43 (23.5\%) of the 183 production packages are not always shipped to production. This indicates that some packages are used differently (in production and not in production) across projects regardless of their functionalities. We show in Table \ref{tab:in_between_packages} 10 examples of such packages, and how often they are in production versus how often they are installed. The results show that \texttt{react}, a library for building user interfaces, is the most frequently installed package appearing in 40 projects, but is only released to production in 4 projects. 
In contrast, \texttt{react-redux}, a React binding for Redux allowing React components to read data from a Redux store, only appears in 5 projects, but is released to production in 4 out of 5 projects. 
In only one project (redux-little-router) is \texttt{react-redux} not released to production and declared as a development dependency. We further inspect the \packagejson of \texttt{redux-little-router} and find that \texttt{react-redux} is a peer dependency, thus, it is not included in the production bundle of the project. 
It is also worth noting that redux-little-router is a lightweight library that provides flexible React bindings and components. Thus, the project makes a conscious effort to mitigate dependency bloat, declaring most of its dependencies as development dependencies, and including \texttt{react}, \texttt{react-dom}, \texttt{react-redux}, and \texttt{redux} as peer dependencies.

The main takeaway from this finding is that we cannot identify production dependencies by looking at the functionalities of a package alone. 
As we have shown, the scope of a dependency can vary based on the context and usage of a package, which means it may differ from project to project. For example, a module bundler may be used in production in one application since it uses some of its functionalities at runtime, but may only be used in development in another application. Thus, it is important for SCA tools to include this scope analysis in their approach so that developers can more easily identify production dependencies based on their own usage and context. 

\begin{table}[t]
\caption{Frequently installed packages that are both used and not used in production. \label{tab:in_between_packages}}
\centering
\small
\begin{tabular}{rrrr}
\toprule
\textbf{Package}  & \multicolumn{1}{c}{\textbf{\# Production}} & \multicolumn{1}{c}{\textbf{Total \# }} & \multicolumn{1}{c}{\textbf{\% in }} \\
                    & \textbf{Installations} & \textbf{Installations} & \textbf{Production} \\
\midrule
\texttt{react}                  & 4            & 40          & 10\%        \\
\texttt{react-dom}            & 3             & 37           & 8.1\%         \\
\texttt{prop-types}                     & 13              & 23            & 56.5\%          \\
\texttt{@babel\/runtime}                        & 10               & 19            & 52.65\%           \\ 
\texttt{lodash}                       & 4               & 14            & 28.6\%           \\ 
\texttt{core-js}                      & 5               & 13            & 38.5\%           \\ 
\texttt{classnames}                       & 5               & 8            & 62.5\%           \\ 
\texttt{react-is}                      & 1               & 5            & 20\%           \\ 
\texttt{react-redux}                    & 4               & 5            & 80\%           \\
\bottomrule
\end{tabular}
\end{table}

\conclusion{Our findings indicate that 28.2\% of production dependencies come from development dependencies and that 51.8\% come from transitive dependencies. The functionality of the package alone does not determine if they will be shipped to production: 43 of 183 packages encountered in production in one project are not shipped to production in other projects. }

\subsection*{RQ3: \rqiii}
\noindent\textbf{Motivation:} 
The observations made in RQ1 suggest that the majority of the dependencies are not used in production. 
While vulnerabilities in non-production dependencies may affect the development environment (e.g., installing packages with malicious code), it is when a vulnerable dependency is released to production that the threat of exploitation reaches its peak \cite{zapata}.
Developers should constantly run scanners to identify security alerts in their project and prioritize fixes in production dependencies, to avoid having their software compromised. The problem is that tools such as \texttt{npm audit} often report many false alerts for deployed code, making vulnerability reports noisy and bloating audit resources \cite{pash, help}. 
In this research question, we investigate how often security alerts are emitted for production dependencies compared to non-production dependencies and the characteristics of vulnerable dependencies.

\noindent\textbf{Approach:} 
To investigate how often vulnerabilities are encountered in production and non-production dependencies, we first generate the \npm vulnerability report of each project by using the \texttt{npm audit} tool. Next, to obtain the \texttt{npm audit} reports in a parseable csv format, we adapt the \texttt{npm-deps-parser} \cite{a2020_npmdepsparser}, a tool that parses, summarizes, and prints \texttt{npm audit} json output to markdown. From this, each vulnerability report is identified with the project name, the vulnerable dependency and version, the severity, and a unique link to the GitHub Advisory Database (GAD) \cite{a2022_github}, a database of
security advisories affecting the open source world. 
To obtain the scope and depth of each vulnerable dependency, we cross-reference the set of vulnerable dependencies with the set of production dependencies and  installed dependencies for each project. 
Because of the limitations discussed in Section \ref{sec:missing_peer}, we could not identify the scope and depth of 29 vulnerable dependencies and exclude them from further analysis. 
\\

\noindent\textbf{Finding 7: A total of 608 security alerts are emitted for dependencies of 32 projects, yet none are related to production dependencies.}
In our dataset, no security alerts were emitted for 68 projects. 
The remaining 32 projects reported a total of 608 security alerts for 456 vulnerable dependencies, i.e., the same dependency may issue multiple security alerts. 
In median, these 32 projects reported 16 security alerts, none related to production dependencies. 

There are a few reasons as to why security alerts may have been emitted only to non-production dependencies. 
First, as seen in RQ1, the vast majority of dependencies are not released to production (99\%), the chances of vulnerabilities being encountered in non-production are 99x higher than in production dependencies.
Second, developers of the selected projects are likely making the conscious effort of updating production dependencies to mitigate security vulnerabilities, since they may be aware of what dependencies may be used in production (e.g., developers open a PR in the project \texttt{InstantSearch} to update a vulnerable dependency~\cite{featdepe77:online}). 
The problem, however, is that tools such as \texttt{npm audit} make no distinction whether security alerts are referring to non-production dependencies.
Developers have to know themselves which dependencies are released to production to filter out relevant security alerts that need urgent action, making it harder to prioritize management efforts. This is shown in related work~\cite{Kula_2017}, where authors reported that 69\% of the surveyed developers claimed to be unaware of their vulnerable dependencies and that dependency updates are perceived as extra workload and responsibility.
\\

\begin{table}
\centering
\caption{Characteristics of vulnerable dependencies reported by \npm vulnerability alerts. \label{tab:vuln_deps_characteristics}}
\small
\begin{tabularx}{.8\linewidth}{rrrr}
\toprule
            & \textbf{Direct} & \textbf{Transitive} & \textbf{Total} \\ \midrule
\textbf{Development} & 9      & 410        & 419   \\ 
\textbf{Runtime}     & 1      & 7          & 8     \\ \midrule
\textbf{Total}       & 10     & 417        & 427  \\
\bottomrule 
\end{tabularx}
\end{table}

\noindent\textbf{Finding 8: 98.1\% (419) of the vulnerable dependencies are development dependencies.} 
In this analysis, we switch from security alert reports to vulnerable dependencies, as multiple reports may be issued for the same dependency under different vulnerabilities. The first row of Table \ref{tab:vuln_deps_characteristics} shows the number of development and runtime vulnerable dependencies reported by our experiment.  
The \texttt{npm audit} tool reports security alerts from a total of 419 vulnerable development dependencies, representing 98.1\% of all vulnerable dependencies identified.
From the 419 vulnerable development dependencies, 9 (2.1\%) are direct dependencies, and 410 (97.9\%) are transitive dependencies. 
Next, we analyze the severity of the vulnerability reports in relation to the characteristics of vulnerable dependencies as shown in Table \ref{tab:severity}. We find that all critical and most of the high-severity reports are emitted for development dependencies.

In some cases, tools such as \texttt{npm audit} allow developers to filter out development dependencies from the security reports, as they are supposedly not released to production \cite{help}. 
It is dangerous, however, to completely ignore the security maintenance of development dependencies. 
Some development dependencies are used in production, as seen in RQ2, and hence, have the risk of being exploited in a production environment. 
Vulnerable transitive dependencies that are released to production are equally dangerous since even if they are reported by \npm, developers are not in control of their update. \\

\begin{table}
\centering
\caption{Count of vulnerability reports per severity level with the \npm recommended action.\label{tab:severity}}
\small
\begin{tabular}{l l|r r}
\toprule

\multicolumn{2}{c|}{\textbf{Vulnerability Severity}} & 
\multicolumn{2}{c}{\textbf{Dependency}} \\

\textbf{Severity} & \textbf{Recommended action} & \textbf{Runtime} & 
\textbf{Dev} \\
\midrule

low                                    & Address at your discretion                       & 3            &   33                  \\
moderate                               & Address as time allows                           & 1           &   226                  \\
high                                   & Address as quickly as possible                   & 5           &   263                \\
critical                               & Address immediately                              & 0            &   45                  \\ 
\bottomrule

\end{tabular}
\vspace{-.4cm}
\end{table}

\conclusion{32 projects in our dataset reported a total of 608 security alerts, but none of the alerts referred to a production dependency. 
Projects have in median 16 security alerts, but the vast majority refer to development non-production dependencies which does not represent a threat for their running application. }

\section{Discussion}
\label{sec:discussion}

In this section, we discuss the implications from our work and possible solutions for current source code based tools.

\subsection{Implications}
\label{sec:current_state}

\textbf{Tracking production dependencies is very challenging.}
While our study focuses on a selected number of 100 popular JavaScript projects, the results showcase the difficulties of mapping a project's production dependencies.
This difficulty arises primarily because assumptions commonly held by the development community regarding dependency management do not hold in practice:

\begin{enumerate}
    \item Assumption 1: Runtime dependencies are always shipped to production. Our results showed that the majority of dependencies declared as runtime are not used in production (RQ1). 
    Developers may spend time ineffectively managing runtime dependencies due to security alerts, without confirming that such dependencies are bundled in their delivered software.

    \item Assumption 2: Development dependencies are never shipped to production.
    In projects that use module bundler, dependencies declared as development may be shipped to production (RQ2). In fact, development dependencies represent a third of all production dependencies identified in our study. 
    Developers may disregard all their development dependencies as being irrelevant for security upgrades, when in fact, some vulnerable development dependencies are shipped in their delivered software.
    
    \item Assumption 3: The functionality provided by the package is sufficient to determine if it is a production dependency.
    Particularly in cases where packages provide runtime utilities, our results show that 43 out of 183 packages are released to production in some projects but not in others.  
    Thus, the package's functionality is not sufficient to determine whether a package is used in production (RQ2).
\end{enumerate}

These assumptions have the potential to affect the security of the delivered software, as developers may wrongly assume what dependencies are sensitive to security exploits. 
\\

\textbf{Not all vulnerabilities in dependencies are a security risk for the software in production.} 
Prior research has shown that not all vulnerabilities are relevant for the software in production \cite{pashchenko, pash}. 
In this paper, we expand on this by studying the relevance of dependencies for the security risk of a software in production. 
Our findings indicate that, given by the prominence of non-production dependencies (RQ1), the vast majority of security alerts will be emitted for dependencies that do not impact the security of a software in production (RQ3).

To put things in perspective, we analyze the types of vulnerabilities reported by \texttt{npm audit}. 
The most common type of vulnerability identified in the studied projects is Regular Expression Denial of Service (ReDoS), accounting for 25.3\% of all reported vulnerabilities and for 27\% of high severity vulnerabilities.  
While a diligent dependency management is of utmost importance to mitigate security risks, developers should be mindful of the types of alerts they should prioritize. 
In the case of a ReDoS attack, the performance of an application is compromised if there is a regular expression that, with malicious input, slows it down exponentially. However, we find that 97.3\% of the dependencies affected by a ReDoS vulnerability are development-only, which tend not to be part of a production software.  
Previous research shows that developers tend to ignore security alerts when they receive a lot of them \cite{Kula_2017}. Our approach allows them to focus on the important ones first (i.e., security alerts for vulnerable dependencies in production).
\\

\noindent
\textbf{Source maps and tree shaking can benefit developers beyond client-side applications.} 
In this paper, we use module bundlers to accurately differentiate between installed dependencies and production dependencies. 
Module bundlers are most commonly used in client-side applications, which are generally defined as libraries or frameworks running in a Web browser (e.g., React, Vue, Angular) to support the development of Web applications. 
Module bundlers, however, can benefit far beyond just client-side applications by helping developers:

\begin{enumerate}
    \item Prioritize addressing security alerts on production dependencies. 
    As security alerts are very commonly issued for projects that rely on open source code, developers should prioritize addressing security issues that have the potential to affect their production software, by identifying vulnerabilities affecting their production dependencies. 
    
    \item Prioritize maintenance tasks on production dependencies. 
    As projects depend on increasingly high number of software dependencies, updating all dependencies in every release may become increasingly prohibitive.
    Updating dependencies always have the risk of breaking changes~\cite{bogart_2021_when}, leading to software bugs and mistrust between project maintainers~\cite{javanjafari_2021_dependency}.
    Hence, developers should prioritize updating production dependencies to focus their maintenance tasks on packages that may affect their delivered software.   
    
\end{enumerate}

\vspace{-.5cm}

\begin{table*}[t]
\small
\vspace{-.4cm}
\caption{Vulnerable dependency (VD) characteristics based metrics reported by current tools and support to locate vulnerable code (VC) in JavaScript projects.\label{tab:tool_comparison}}
\centering
\begin{tabular}{r|c|ccccc|c}
\toprule

    & \textbf{Cover all} & \multicolumn{5}{c|}{\textbf{Filter Security Alerts by}} & \textbf{Locate} \\

\textbf{Tool} & 
\textbf{dependencies} & 
\textbf{Runtime} & 
\textbf{Development} &
\textbf{Direct} &
\textbf{Transitive} & 
\textbf{In production} &
\textbf{dep code} \\

\midrule

\textbf{npm audit}       
    & Yes
    & Yes       
    & Yes
    & Yes
    & Yes
    & No          
    & No        
    \\
    
\textbf{Snyk}    
    & Yes
    & Yes       
    & Yes
    & Yes
    & Yes         
    & No           
    & No         
    \\

\textbf{Dependabot} 
    & Yes
    & No          
    & No
    & No
    & No
    & No            
    & No          \\
    
\textbf{OSWAP Dependency-Check}       
    & Yes
    & No          
    & No
    & No
    & No             
    & No            
    & No           \\ 
    
    \bottomrule
\end{tabular}
\vspace{-.2cm}
\end{table*}

\subsection{Towards Better Tool Support}
\label{sec:better_support}

SCA tools are constantly used by software projects to control the risk related to software dependencies, such as vulnerabilities, and compliance to open source licenses\cite{sca_def}.
To understand the support current SCA tools provide to production dependencies, we investigate four popular tools: npm audit, Snyk, Dependabot, and OSWAP Dependency-Check. 
We analyze the documentation of the SCA tools, as well as apply them to some of our studied projects to assess their capabilities and limitations.

We present in Table \ref{tab:tool_comparison} an overview of the features related to dependency scope and usage from four popular SCA tools. 
All SCA tools we assess cover all dependencies of a software project, including both direct and transitive dependencies.  
Given a project may have thousands of installed dependencies, we now dive into the filtering capabilities of the tools. 
We note that only \texttt{npm audit} and Snyk~\cite{snyk} provide ways of filtering security alerts based on whether the vulnerability affects runtime/development dependencies or direct/transitive dependencies. 
The filtering of runtime/development dependencies is based on project configuration (e.g., \packagejson file), thus, it is subject to limitations when it comes identifying production versus non-production dependencies. 
Neither Dependabot nor OSWAP Dependency-Check allow users to filter security alerts based on the scope or depth of their dependencies.

It is worth noting that none of the tools provide a way to differentiate between production versus non-production dependencies. 
There is no support, for instance, to input source map files in the tools, to help filter out vulnerabilities that concern non-production dependencies. 
We believe adding source map support to SCA tools would offer developers better insight on their production bundle without relying so much on the dependency configurations that have shown to be inconsistent across different projects.

Finally, we find that none of the tools provide a way to locate where the vulnerable dependency is used in code (column ``Locate dep code''). 
Developers have to rely on their own set of static/dynamic analysis tools to know exactly where the vulnerable dependency is used in the codebase. 
We believe  static analysis tools would benefit from using the features provided by module bundlers that scan the code for \texttt{import} statements to provide the path to the source file in which a dependency is imported and used.

\section{Related Work}
\label{sec:related_work}

In this section, we discuss the related literature divided into three aspects. First, we discuss works that have focused on the challenges related to the Software Bill of Materials. Then, we discuss works describing the challenges of dependency management in software ecosystems. Finally, we discuss existing tools and approaches to detect vulnerable dependencies.

\subsection{Software Bill of Materials}
\label{sec:bom}

The Cybersecurity and Infrastructure Security Agency (CISA) defines the Bill of Materials (BOM) as a nested inventory of components in a piece of software \cite{software}. The process of identifying production dependencies is part of the constructing the BOM of a software. Several studies have proposed approaches to consolidate the BOM of software applications \cite{abdalkareem_2017_why, Costa_TSE2021, mujahid_2021_toward}. Zajdel et al. discussed that users of open source softwares tend to arbitrarily download the software into their build systems, but rarely keep track of which versions they use which results in unnecessary software being left in the application, increasing the risk of potential vulnerabilities \cite{Zajdel:22}. Coelho et al. proposed a data-driven approach to measure the level of maintenance activity in GitHub projects \cite{DBLP:journals/corr/abs-2003-04755}. The authors found that 16\% of the studied open source projects have become unmaintained over the course of one year. They also reported that software tools such as compilers and editors have the highest maintenance activity over time and proposed that a metric about the level of maintenance activity of GitHub projects can help developers in selecting open source projects. 

These prior studies focus on the importance of selecting well-maintained software libraries, and propose approaches to alleviate the challenges related to open source code reuse. However, the challenges of constructing the BOM for dynamic languages like JavaScript are still present. Thus, our study focuses on JavaScript projects and leverages existing approaches (i.e., module bundlers and tree shaking) to help developers maintain their softwares and decrease the risks related to open source code reuse by analyzing the software's dependencies and reporting on the ones that are actually used in the code.

\subsection{Dependency Studies} 
\label{sec:dep_studies}

Package ecosystems and the presence of vulnerable dependencies have been studied in the literature \cite{javanjafari_2021_dependency, decan, abdalkareem_2020_on, kikas}. Hejderup et al. report that one-third of the \npm packages use vulnerable dependencies \cite{hejderup_2015_in}. Similar to our study, the authors suggest context of usage of a package to be a possible reason for not fixing the vulnerable dependencies. Abdalkareem et al. conduct an empirical analysis on security vulnerabilities in Python packages \cite{abdalkareem_2020_on}. They find that the number of vulnerabilities in the \texttt{PyPi} ecosystem increases over time and that it takes, on median, more than 3 years to get discovered, regardless of their severity. They emphasize on the need for more effective process to detect vulnerabilities in open source packages since both \npm and \texttt{PyPi} allows to publish a package release to the registry with no security checks. Lauinger et al. conduct the first large scale of JavaScript open source projects and investigate the relationship between outdated dependencies and dependencies with known vulnerabilities \cite{Lauinger_2017}. They report that transitive dependencies are more likely to be vulnerable since developers may not be aware of them and have less control over them, which further corroborates with our findings of RQ3. Similarly, Williams et al. report that 26\% of open source Maven packages have known vulnerabilities and refer to a lack of meaningful controls of the components used in the proprietary projects as a partial explanation to this high number of vulnerable dependencies \cite{williams2012unfortunate}. 

These prior studies focus on the presence of known vulnerabilities in popular package ecosystems and the reason why the number of vulnerable dependencies is so high. However, their analysis does not consider the scope of dependencies (i.e., they do not distinguish production and non-production dependencies). As a result, the studied vulnerable dependencies may not be exploitable. Zapata et al. investigate vulnerable dependency migrations of \npm packages and evaluate the impact of a vulnerability in the \texttt{ws} package on 60 JavaScript projects using the vulnerable version of the package \cite{zapata}. The authors find that up to 73.3\% of the dependent applications were safe from the vulnerability since they did not actually used the vulnerable code. The study also highlights that it is not trivial to map vulnerable code to client usage for JavaScript, which further corroborates with our findings in RQ1.

\subsection{Detecting Vulnerable Dependencies} 
\label{sec:related_vulnerabilities}
Alfadel et al. study the use of Dependabot security pull requests in 2,904 JavaScript open source GitHub projects \cite{alfadel_dependabot}. Results show that the vast majority (65.42\%) of the security-related pull requests are often merged within a day and that the severity of the vulnerable dependency or potential risk for breaking changes are not associated with the merge time. Ponta et al. propose a pragmatic approach to facilitate the assessment of vulnerable dependencies in open source libraries by mapping patch-based changes of vulnerabilities onto the affected components of the application \cite{ponta}. Sejfia et al. present \texttt{Amalfi}, a machine-learning based approach for automatically detecting potentially malicious packages \cite{sejfia2022practical}. The authors evaluate their approach on 96,287 \npm package versions published over the course of one week and identify 95 previously unknown vulnerabilities. Pashchenko et al. propose Vuln4Real, a methodology that addresses the over-inflation problem of academic and industrial approach for reporting vulnerable dependencies in free and open source software (FOSS) \cite{pashchenko}. Vuln4Real extends state-of-the-art approaches to analyzing dependencies by filtering development-only dependencies, grouping dependencies by project, and assessing dead dependencies. Their evaluation of Vuln4Real shows that the methodology significantly reduces the number of false alerts for code in production (i.e., dependencies wrongly flagged as vulnerable). Pashchenko et al's. work is the closest to ours since it considers similar aspects in relationship to the relevance of vulnerable dependencies: exploitability and dependency scope. Our study touches on another aspect that is not discussed in Vuln4Real and that is the context in which a dependency is used versus how it is configured. Our paper shows that there is a discrepancy between the configuration of dependencies and its usage, and that this discrepancy may affect the exploitability of a vulnerability (i.e., its relevance to the application). 

Imtiaz et al. \cite{imtiaz_2021_a} present an in-depth case study by comparing the analysis reports of 9 SCA tools on OpenMRS, a large web application composed of Maven and \npm projects. The study shows that the tools vary in their vulnerability reporting and that the count of vulnerable dependencies reported for \npm projects ranges from 32 to 239. From the 9 studied SCA tools, 4 freely available tools could be applied to \npm projects: OWASP Dependency-Check, Snyk, Dependabot, and npm audit. The results show that all 4 tools detect vulnerable dependencies across all scopes and depths and that reported vulnerabilities are mostly introduced through transitive dependencies, except for Dependabot. While the authors of this paper report on the coverage capabilities of SCA tools, our study mainly focuses on the data that is shown to the user. For example, npm audit covers dependencies of all scope when reporting for vulnerabilities, but it is the user's responsibility to filter the vulnerable dependencies by scope (production or development). That is, SCA tools don't explicitly report on the scope of vulnerable dependencies, and when it is done manually by users, this analysis depends on the project's dependency configurations rather than dependency usage.

\section{Threats to Validity}
\label{sec:threats_to_validity}

\noindent
\textbf{Threats to internal validity} considers the experimenter's bias and errors.
Our method of analysis relies on building software projects with their configured module bundler to identify production dependencies, and errors in this process may introduce false positives/negatives in our analysis. 
We mitigate this threat by 
1) only selecting projects that already use module bundlers to minimize any intervention that could introduce bugs in the process, 
2) building each project manually by following the projects documentation, 
3) manually inspecting the built artifacts (e.g., installed dependencies, source map files), and
4) removing 55 projects that showed evidence of failed builds (e.g., errors, empty source map files).
To further confirm the validity of this process, we also sampled 7 projects from our dataset and asked contributors to validate our results, by checking the accuracy of the yielded production dependencies. 
We received responses from 4 projects and contributors confirmed the yielded classifications, helping us validate the soundness of our methodology.

\textbf{Threats to eternal validity} considers the generalizability of the findings. 
We purposefully select projects that already use module bundlers which could limit the type of project our findings generalize. 
First, module bundlers tend to be used primarily by projects that want to minimize their production dependencies, such as client-side packages such as web-applications and libraries.
In fact, our finding that development dependencies are shipped to production are unlikely to occur in projects that do not use module bundlers. 
Second, our dataset is strictly composed of open source JavaScript projects, thus, our results may differ if a study is performed on proprietary projects or projects written in other languages.

\section{Conclusion and Future Work}
\label{sec:conclusion} 
This research investigates projects dependencies that are released to production and their impact on security and dependency management. We conducted our study on 100 \npm projects, one of the largest and fastest growing software ecosystems. 
Our results showed that production dependencies are rare among the installed dependencies of a project, but are difficult to identify. 
Commonly held assumptions of dependency management do not hold in practice and context is more important in determining the scope of a dependency as opposed to its configuration.
Furthermore, we evaluate how often security alerts are reported for production dependencies, and found that none of the vulnerability reports are emitted for dependencies released to production. Rather, the majority of the alerts are emitted for development, transitive dependencies which has two main implications: 1) not every vulnerability is a threat to the software in production, and 2) vulnerabilities can be introduced transitively regardless of their scope, which further motivates the need for SCA tools to provide such an analysis.

Our paper outlines directions for future work. 
Using module bundlers as a way to identify production dependencies may augment current SCA tools to provide better insights on the scope of their dependencies within their project's context and usage.
Consequently, module bundlers or similar tools, may benefit far more than just client-side applications and should be part of the build process of projects that extensively rely on open source code.

\bibliographystyle{ACM-Reference-Format}
\bibliography{main.bib}

\appendix

\end{document}